\documentclass[12pt]{article}
 \usepackage{graphics}
\usepackage{graphicx}

\usepackage{amssymb}
\title{Initial data for general relativistic SPH with Centroidal Voronoi Tessellations}
\author{Juan Pablo Cruz P\'erez, Jos\'e Antonio Gonz\'alez C.}
\begin{document}
\maketitle


\begin{abstract}
In this work we present an alternative method to obtain a distribution of particles over
an hyper surface, such that they obey a rest-mass density distribution $\rho(x^i)$. 
We use density profiles that can 
be written as $\rho(x^1,x^2,x^3)=\rho(x^1) \rho(x^2) \rho(x^3)$ in order to be able to 
use them as a probability density functions.  We can find the relation between the chart $x^j$ and
a uniform random variable $\bar{x}^j \in (0,1)$, say $F(x^j)=\bar{x}^j$. 
Using the inverse of this function we relate a set of $N$ arbitrary number of
points inside a cube with coordinates $\{ x^j =F^{-1}(\bar{x}^j)\}$ giving the position in order to
get the density distribution $\rho(x^j)$. We get some noise due to the
random distribution and we can notice that each time we relax the configuration on the cube
 we also get a better distribution of the desired
physical configuration described with $\rho(x^j)$. This relaxation of the position of the particles
in the cube has been performed a Lloyd's algorithm in 3D and we have used {\it Voro++} library in order to get the Voronoi tessellations.
\end{abstract}


\section{Introduction}
\label{sec:Intro}
In Smoothed Particles Hydrodynamics (SPH), the initial configuration of the free mesh is done thinking
on the nodes as particles with equal mass. However, this represents a difficult task because we need to
set up certain number of particles such that they reproduce a density profile using SPH averages.
The methods for generating the initial distribution of the points
can be done using glass-like configurations similar to the used for the rigid grid configurations, this kind of configurations
are easily computed for uniform density profiles however for more complicated density profiles the
position of the particles are not easy to determine because we are constrained by the condition of
particles having the same mass \cite{SRosswog}\cite{Jprice_magneto}. 
Monte Carlo Method's (MC), also known as acceptance rejection and coordinate transformation\cite{numrec}, are the most popular for building this
kind of initial configuration with equal mass particles.
Despite the relative error obtained with this kind of methods is of order $O(1/\sqrt{N})$ the condition of 
equal mass particles is compatible with $MC$ methods. However the error produced by the 
random sorting of particles will corrupt enormously the numerical evolution. For this reason
people using SPH codes have used relaxed configurations of random distributions or cristal kind configurations
if it is possible (for uniform distributions).\cite{voro_configurations}
In this work we present an algorithm which can be used for generating a distribution of $N$ particles
spread over a curved space such that each one of the particles posses the same rest-mass 
obeying  a density profile with spherical symmetry provided analytically or numerically. This method can 
be used in newtonian distributions or general(special) relativistic initial configurations. 
Using basic elements of statistics and probability theory we can establish a relation between a three dimension
uniform distribution of particles with an spherical distribution in flat or curved space.
 Our claim is the following: if we get a better distribution of particles
for a uniform distribution this should be inherited to the physical configuration using a MC method, specifically 
the coordinate transformation.

The content of this paper is organised as follows: in the section \ref{sec:coord_trans} we describe the transformation
of coordinates, this method describes the correct way for generation of particles distributions in curved space with a desired
profile, we assume spherical symmetry.
In section \ref{sec:sph} we present the basic approximation of the SPH method.
In section \ref{sec:voro} introduce the Voronoi tessellation and the relaxation method in the uniform distribution
using the Lloyd's algorithm, in section {\ref{sec:method}} we describe our method and in section  \ref{sec:applications} we present
two examples: a uniform density profile in a flat space-time, and an spherical 
distribution of mass on a regular curved space-time known as Tolman-Oppenheimer-Volkoff (TOV)\cite{Holes-Dwarfs-Shapiro}.
\section{Coordinate Transformation}
\label{sec:coord_trans}
In order to study the evolution of any physical system we need to formulate it as a Cauchy problem.
Hydrodynamical evolutions in (fixed) space-time can be treated in this way using $3+1$-formalism.
Writing the metric in $3+1$-form we foliate the space-time into a family of three dimensional space like 
hyper surfaces of constant $t$, $\Sigma_t$. The line element is given by 
\begin{equation}
ds^2 = g_{\mu \nu}dx^\mu dx^\nu= -(\alpha -\beta^2 )dt^2 +2 \beta_i dx^i dt + \eta_{ij} dx^i dx^j,
\end{equation}
\noindent where $\alpha$ is the lapse function, $\beta^i$ is the shift vector and $\eta_{ij}$ is 
the 3-metric induced on the spacial hyper surface  $\Sigma_t$, for a 
coordinate system ${t,x^j}$. 
Using the Gauss theorem over the baryon conservation law we can find the rest mass contained on a region $V\subset \Sigma_t$
\begin{equation}\label{eq:mo_rel}
M_o = \int_{V\subset \Sigma_t} \rho(x^j) u^t \alpha \sqrt{\eta} d^3 x,
\end{equation}
\noindent here $\rho$ is the rest mass density, $u^t$ is the $t-$component of the four velocity
and $\eta$ is the determinant of the induced metric on $\Sigma_t$. For a detailed description
for $ADM$ see \cite{nr_miguel} \cite{NR_baumgarte}. 
We are going to relate $M_o$ with the concept of probability
using some basic concepts \cite{prob_book}.

{\bf Probability distribution/probability density function:} Let $X$ be a continuous random variable, then a
{\it probability distribution} or {\it probability density function} (p.d.f) of $X$ is a function $f(x)$ such that 
for any two numbers $a$ and $b$ with $a \leq b$,
\begin{equation}\label{eq:prob_def}
P(a\leq X\leq b) = \int^b_a f(x) dx.
\end{equation}
That is, the probability that $X$ takes on a value in the interval $[a,b]$ is the area under the graph of the
{\it density function}.
The (p.d.f.) need to satisfy two conditions in order to be legitimate:
\begin{itemize}
\item {{\bf(i)} $f(x) \ge 0$ for all $x$,}
\item {{\bf(ii)} $\int^{\infty}_{-\infty} f(x) dx = 1$, i.e. it need to be normalised.}
\end{itemize}
\noindent Comparing (\ref{eq:mo_rel}) and (\ref{eq:prob_def}) we deduce that the (p.d.f.) will be composed
by the integrand of (\ref{eq:mo_rel}).

{\bf Cumulative distribution function:} The cumulative distribution function (c.d.f.) $F(x)$ for a continuous random
variable $X$ is defined for every number $x$
\begin{equation}
F(x)=P(X\le x) = \int^x_{-\infty} f(y) dy.
\end{equation} 

\noindent where the (p.d.f) comes up as $f_r(r)={4 \pi \rho(r)}/{M_o(R)}$.

{\bf Transformation method:} A computer can generate pseudo-random numbers obeying a 
uniform random distribution. The probability of finding a number in the interval $(x,x+dx)$,
denoted by $p(x)dx$ is given by 
\begin{equation}
 p(x) = \left\{ 
  \begin{array}{l l}
    1 & \textrm{for $0<x<1$}\\
    0 & \textrm{other wise}
  \end{array} \right.
\end{equation}
Suppose that we generate another random variable as a function of the uniform variable $x$, 
lets call it $\Psi(x)$. The probability distribution of $\Psi$, $p(\Psi)d\Psi$ is determined by the 
{\it fundamental law of probabilities}
\begin{equation}
p(\Psi)d\Psi = p(x) dx.
\end{equation}
As an example we can see for an spherical symmetry case, 
in flat space-time the integral (\ref{eq:mo_rel}) is reduced to
\begin{equation}
M_o(R)=\int^R_0 4 \pi \rho(r) r^2 dr \Rightarrow \int^R_0 \frac{4 \pi \rho(r)}{M_o(R)} r^2 dr = 1,
\end{equation}
\noindent where $R$ is the radius of the configuration, $\rho(r)$ is the density profile.
In the case mentioned in the past definition we can see that the (c.d.f) is 
\begin{equation}
F_r(r)= \int^r_{0} f_r(r) dy.
\end{equation} 
Using the (c.d.f) and the Transformation method we obtain
\begin{equation}
\int^r_0f_r(r) dr=:F_r(r)=x=\int^x_0 p(x) dx \Rightarrow r=F^{-1}(x),
\end{equation}
\noindent suppose for simplicity $\rho(r)={M_o(R)}/{2 \pi R^2 r}$, then
\begin{equation}
\left(\frac{r}{R}\right)^{2}=x \Rightarrow r=R\sqrt{x}.
\end{equation} 
In general we have at least for our physical systems more than one coordinates. 
We use the following definition of a multivariate random variables.

{\bf Joint probability marginal function:} For more than two random variables we can define the joint probability 
marginal function of the variables of this function 
\begin{equation}
p(x_1,x_2,\dots,x_n)= P(X_1=x_1, X_2=x_2, \dots, X_n=x_n).
\end{equation}
If the variables are continuous, the joint p.d.f of $X_1,X_2,\dots,X_n$ is the function $f(x_1,x_2, \dots,x_n)$ such
that for any $n$ intervals $[a_1,b_1],\dots,[a_n,b_n]$,
\begin{equation}
P(a_1 \le X_1 \le b_1, \dots, a_n \le X_n \le b_n)= \int^{b_1}_{a_1} \cdots \int^{b_n}_{a_n} f(x_1,\dots,x_n) dx_n \cdots dx_1.
\end{equation}
\noindent Also the transformations method can be generalised to $n$ random variable cases, but if we can define 
$f(x_1,\dots,x_n):=f_1(x_1) \dots f_n(x_n)$, the transformation method can be applied to each variable. 
\begin{equation}\label{eq:joint_separated}
P(a_1 \le X_1 \le b_1, \dots, a_n \le X_n \le b_n)= \int^{b_1}_{a_1}f_1(x_1) dx_1 \cdots \int^{b_n}_{a_n} f_n(x_n) dx_n.
\end{equation}
We have used this property in order to separate in three (c.d.f) and (p.d.f.) each one of the coordinates of
the spherical symmetry distribution.
Using (\ref{eq:mo_rel}) as well as the {\bf p.d.f.(ii)} in order to define  the 
probability function we get the identity
\begin{equation}
1=  \int_{V\subset \Sigma_t} \frac{\rho(x^j)}{M_o} u^t \alpha \sqrt{\eta} d^3 x.
\end{equation}
Using this fact we can define the joint probability function as  the integrand of the integral above
\begin{equation}\label{eq:dpf}
f_{x^1,x^2,x^3}(x^1,x^2,x^3):= \frac{\rho(x^j)}{M_o} u^t \alpha \sqrt{\eta}.
\end{equation}
If the coordinate system and the properties of the density are such that 
$\rho(x^j):=\rho_1(x^1)\rho_2(x^2)\rho_3(x^3)$ and $u^t \alpha \sqrt{\eta}:=g_1(x^1) g_2(x^2) g_3(x^3)$ we can
split (\ref{eq:dpf})
\begin{equation}\label{eq:dpf_decomposed}
f_{x^1,x^2,x^3}(x^1,x^2,x^3):= \frac{1}{M_o} [\rho_1(x^1) g_1(x^1)] [\rho_2(x^2) g_2(x^2)] [\rho_3(x^2) g_3(x^2)], 
\end{equation}
\noindent which is the analogous of the integral given in eq.(\ref{eq:joint_separated}). This 
means that the election of the components does not depend on each other. 
Once we have separated the (p.d.f.) we can deal with one problem for each coordinate 
defining the cumulative function for  $f_{x^j}(x)=\rho_j(x) g_j(x)$  with  $j=1,2,3$ as
\begin{equation}\label{eq:transform}
F_{x^j}(y)=\int^y_0 f_{x^j}(x^j) dx^j.
\end{equation}
\noindent Here $x_j \in D_j$ and $D_j$ is the domain of the definition of the
coordinate. The coordinate transformation can be reached performing 
\begin{equation}\label{eq:inverse}
x^j=F^{-1}_{x^j}(x).
\end{equation}

For a flat space time we get for an spherical distribution $\rho(r,\theta,\phi)=1$, then mass inside an sphere
of radius $R_d$ is given by
\begin{equation}
\int^{R_d}_0r^2dr\int^{2\pi}_0 \sin \theta d\theta \int^\pi_0 d\phi= \frac{4 \pi {R_d}^3}{3},
\end{equation}
From the definition of the cumulative functions we obtain
\begin{eqnarray}
F_r(R)=\int^R_0\frac{3 r^2}{R_D^3}dr=\left(\frac{R}{R_d}\right)^3, \
F_\theta(\theta)=\int^{\theta}_0 \frac{1}{2}\sin \theta d\theta =\frac{1 -\cos(\theta)}{2}, \
F_\phi(\phi)=\int^\phi_0 \frac{1}{ \pi}d\phi= \frac{\phi}{\pi},
\end{eqnarray}
\noindent notice all the cumulative function are normalised.

Then we can use the equation (\ref{eq:transform}) in order to get the transformation between a uniform
distribution and the spherical co-ordinate system $(r,\theta,\phi)$.
$$F_r(R)=x \Rightarrow R=R_d x^{1/3}, \,\,F_r(R)=y \Rightarrow \theta=\cos^{-1} (1-2y),\,\, F_\phi(\phi)=z \Rightarrow \phi=z\pi,$$
\noindent where $x,y$ and $z$ are number provided by a uniform distribution.
	
This algorithm is better than the acceptance rejection method, because it avoids the
waste of random numbers using only the exact amount of calls to the random number
generating subroutine. We spend $(d \cdot N)$ random numbers,  where $d$ is the dimension of the integral.
We call the set of points in the $d$-dimensional space as $A^{(0)}$.

Smoothed Particles Hydrodynamics(SPH) uses Monte Carlo's Theory in order to get the discretised
version of the right hand side of the hydrodynamics equation, this matches with the
initial data generated with the method described above. The following section describes
how to get the standard approximations of SPH.

\section{SPH approximations}
\label{sec:sph}

The Smoothed Particle Hydrodynamics methods are based in two approximations, the 
integral approximation and the particle approximation. The first one is made using the
Kernel function $W$ in order to approximate de Dirac's delta in the identity 

\begin{equation}
f(\vec{x}_a)=\int_{\Gamma}f(\vec{x}') \delta(\vec{x}_a-\vec{x}',h) dV = \int_{\Omega(h) \subset \Gamma} f(\vec{x}') W(\vec{x}_a-\vec{x}',h) dV + O(h^2),
\end{equation} 

\noindent here $\Gamma$ is the region of the space where the distribution of matter is located,
$\Omega(h)$ is the region where the Kernel function is different of zero ($W$ is of compact support),
$dV=\sqrt{\eta}d^3 x$ is the volume invariant.
The second approximation is the change of the integral by a summation over the number of particles
generated using a Montecarlo's method, then using the Benz\cite{sph_laguna} approximation we
obtain

\begin{equation}
\left<f\right>_a = \sum_b \frac{f_b}{n_b} W_{ab}.
\end{equation}

We can recover to the classical SPH approximation considering the flat space $\sqrt{\eta}=1$, $\left| \vec{v}\right| << c$
where $\vec{v}$ is the velocity of the fluid given as an initial data and $c$ is the speed of light.

Using this approximation we can define the relative error particle to particle using

\begin{equation}
e_a:=f(\vec{x}_a)-\left<f\right>_a,
\end{equation}
\noindent from the Monte Carlo's Theory we know that the error is of order $O(1/\sqrt{N})$, and 
the insight of our method consist in the use of a relaxed version of the initial guess $A^o$
using  Lloyd's Algorithm in order to redistribute the particles and to obtain a more evenly distributed
set of particles $A^k$ where $k$ is the number of iterations performed in order to get an accurate
configuration. In the next section we are going to introduce the notion of the Voronoi tessellations and
the detailed description of the Lloyd's algorithm.

\section{Voronoi Tessellations and Lloyd's Algorithm}
\label{sec:voro}

Let $E$ be the set of $N$ points distributed over a region $V \subset R^3$ endowed with an
euclidean metric $d$. For any point $X_k$ we can define the {\it Voronoi cell} as 

\begin{equation}
V_k:= \left\{ x \in V | d(X_j,x) \le d(X_k,x) \forall j \neq k \right\},
\end{equation}

\noindent let us name $E$ as the seed of the Voronoi cells. The center of mass of each
cell is called the centroid or center of mass, which in general is not the same as the generator of cell. 
A {\it Voronoi Tessellation} (VT) could be defined as the set of boundaries of all the 
Voronoi cells, and a {\it Centroidal Voronoi Tessellation} (CVT) is a particular case of VT's when the
center of mass coincides with the Voronoi generator.

The Lloyd's algorithm is an iterative method such that  it gets a CVT obeying a density distribution
$\rho$ starting from an initial guess $A^0$. At each iteration the particle is translated from
the current position to a new one near to the new centroid of the VT. Repeating this 
process until the position of each one of the particles is close as possible to the
centroid of the current VT. 

We generate $A^0$ with $N$ random numbers obeying a uniform distribution for 
three random variables $X^j$. It will occupy a cube 
$C=(0,1)\times (0,1)\times (0,1)$. 
In order the obtain the relaxed configuration we 
use the library {\it Voro++}\cite{chrisvoro++} in order to get the centroids of the VT at $A^0$ and then
we move each one of the particles using
$$X^{j,(1)}=X^{j,(0)}+\lambda(X^{j,(0)}_c-X^{j,(0)}),$$
\noindent where the subscript refers to the centroids. The same process is repeated iteratively until 
$||X^{j,(k+1)}-X^{j,(k)}|| \leq TOL$ for all $j=1,\dots,N$. Here
$TOL$ is defined by the approximated zero of the computer and the kind of data that
we are using. In other words this process
finishes if the configuration does not change anymore. We name to this iteration number:
{\it coldest iteration} and we refer to the iteration number as $k_c$.

A CVT configuration from a uniform density $\rho$ offers some advantages in theory because it will generate
an isotropic configuration of $N$ particles. This fact will
help to perform a better SPH average because every particle would posses the same number
of neighbours distributed evenly around it. We will see in the next section that the coordinate
transformation inherit this property to the desired configuration on the curved space.

\begin{figure}[h!]
  \centering
  \begin{tabular}{cc}
    \includegraphics[height=0.45\textwidth,width=0.5\textwidth]{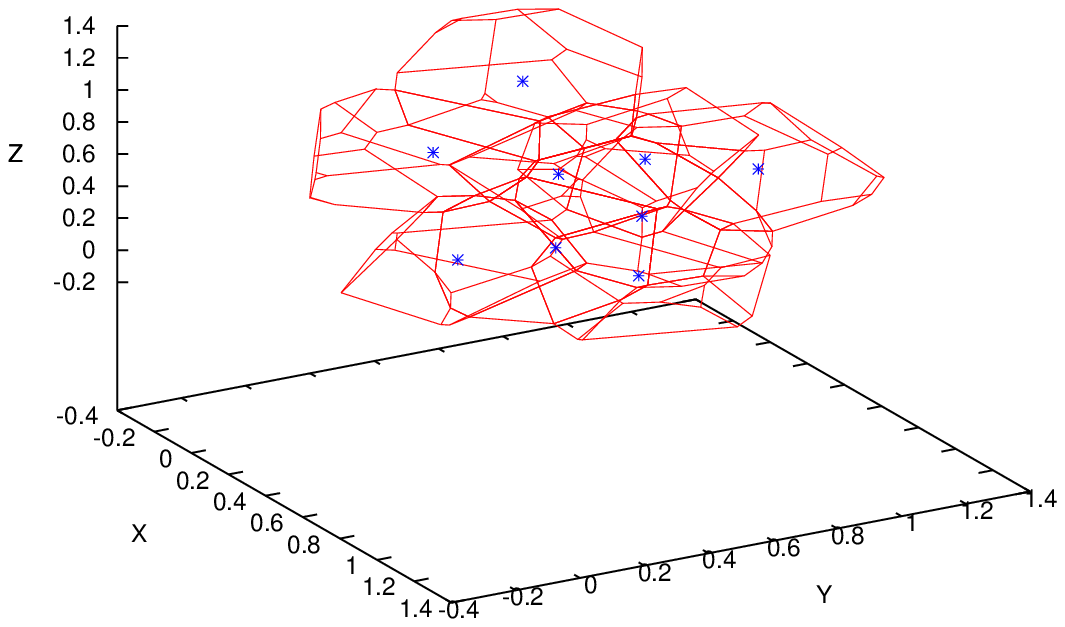}&
      \includegraphics[height=0.45\textwidth,width=0.5\textwidth]{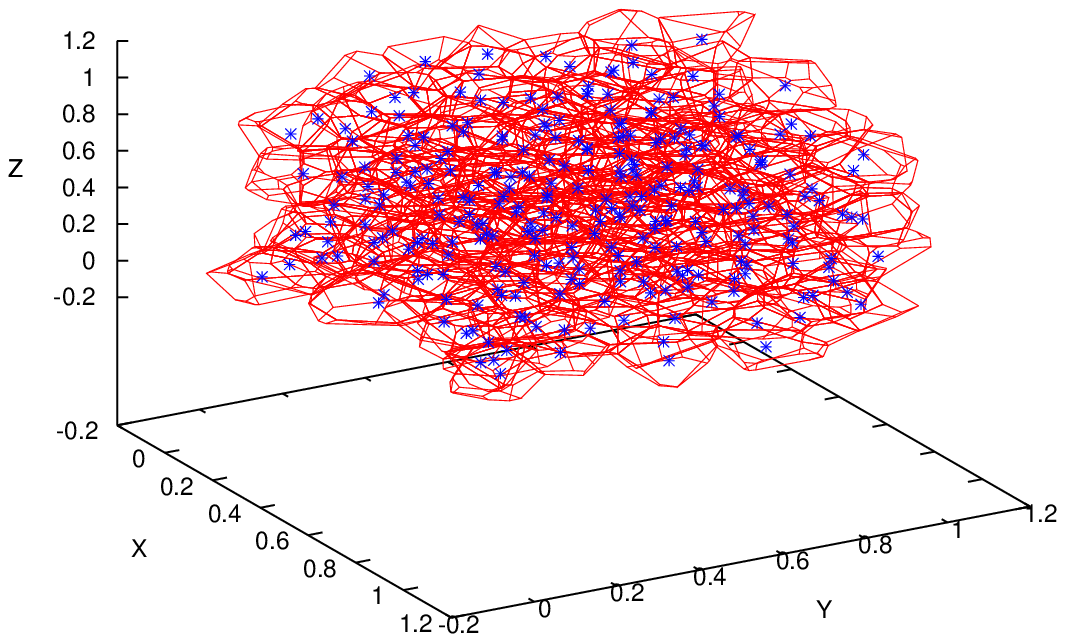} \\     
            \includegraphics[height=0.45\textwidth,width=0.5\textwidth]{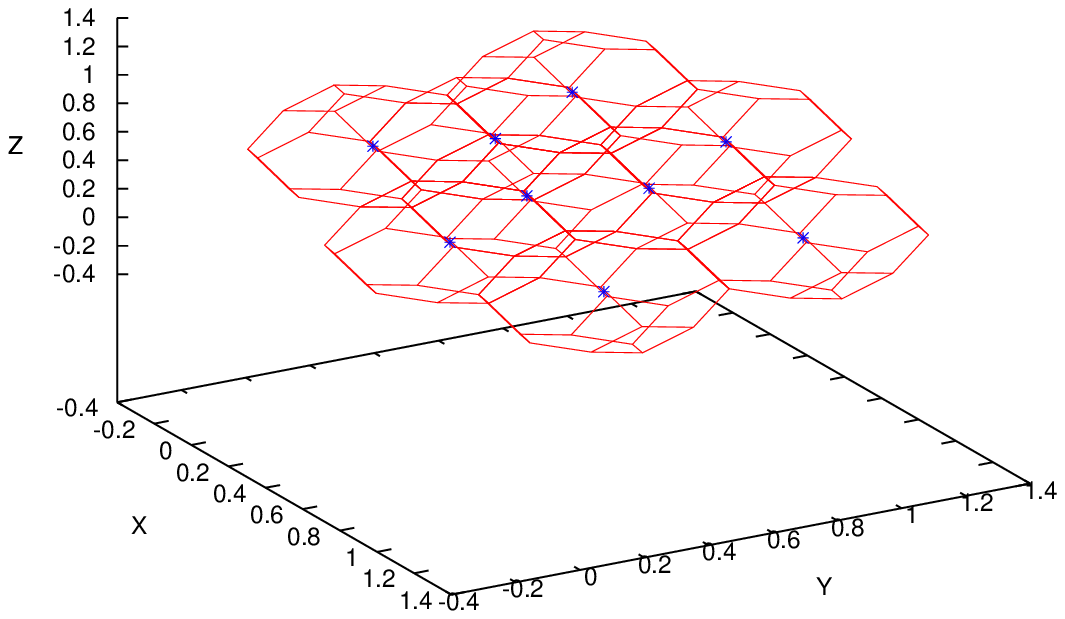}&
             \includegraphics[height=0.45\textwidth,width=0.5\textwidth]{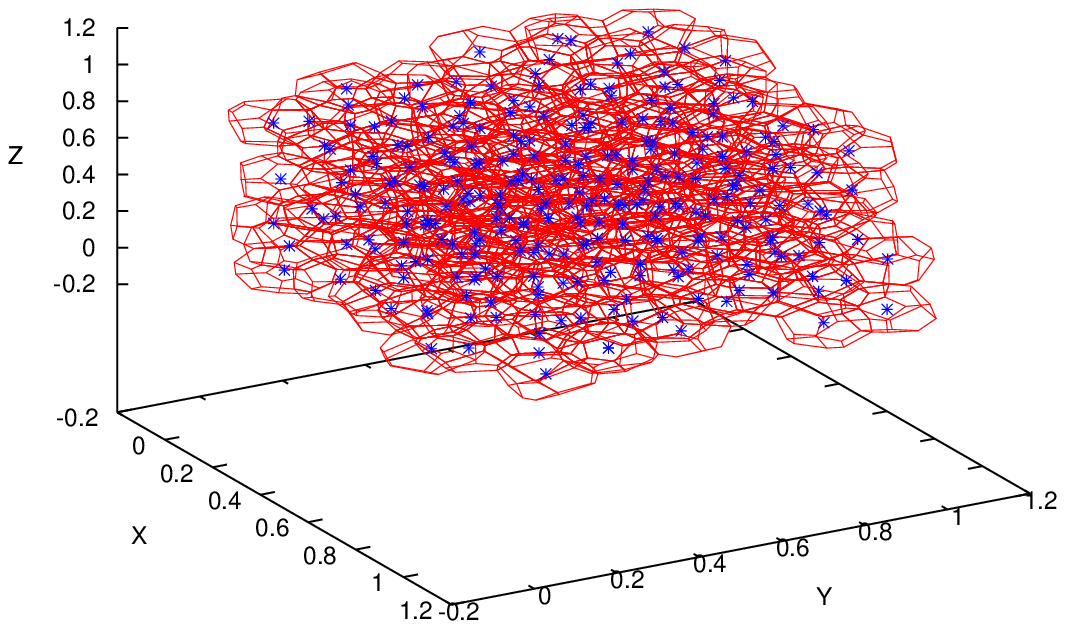}
  \end{tabular}
    \caption{In this plots we show the first and the last iteration for $N=9$(left column)
   and $N=288$(right column). In the left column we can notice the geometric pattern at
   the end of the iterations, but in the right column at the end of the iteration we have 
   a centrical voronoi tessellations but its difficult to notice a pattern.}\label{fig:voro_show}
\end{figure}

\begin{figure}[h!]
  \centering
  \begin{tabular}{cc}
    \includegraphics[height=0.45\textwidth,width=0.45\textwidth]{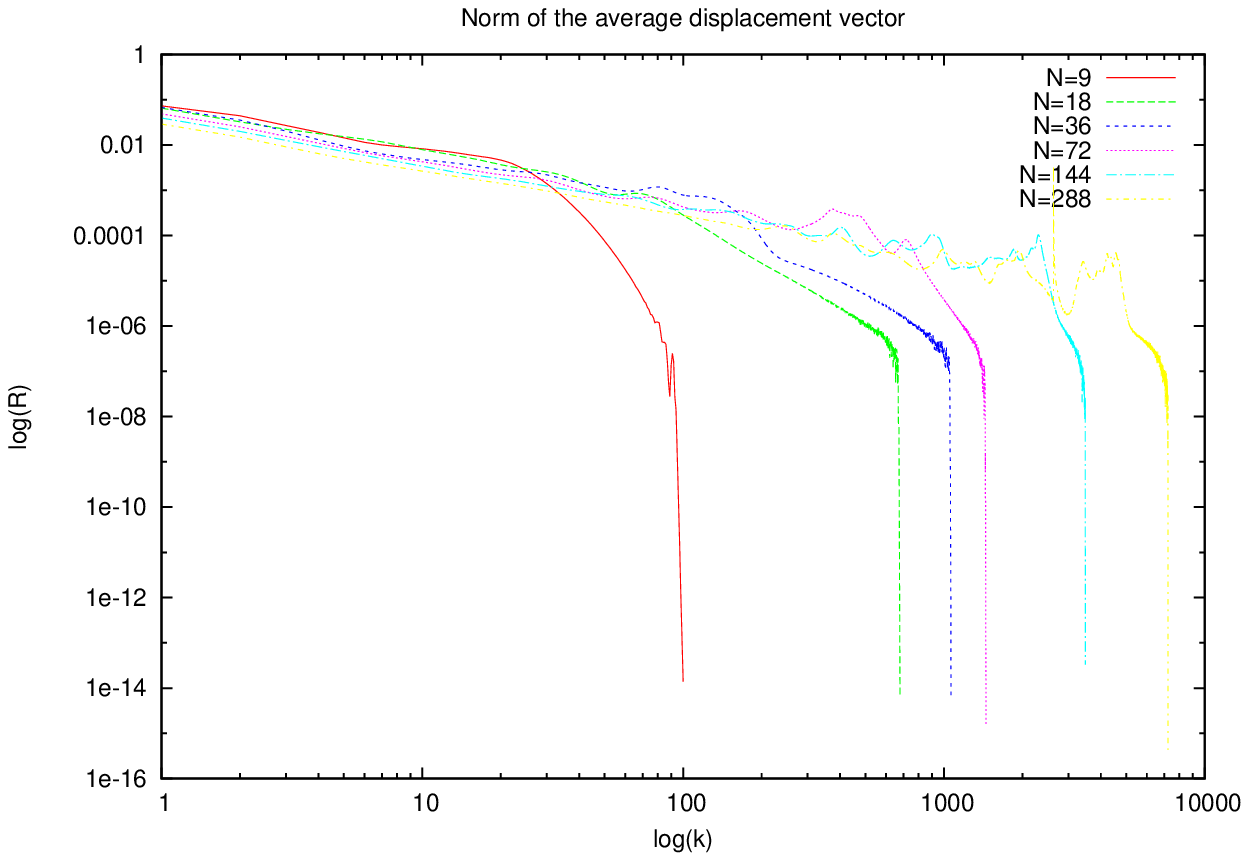}&
      \includegraphics[height=0.45\textwidth,width=0.45\textwidth]{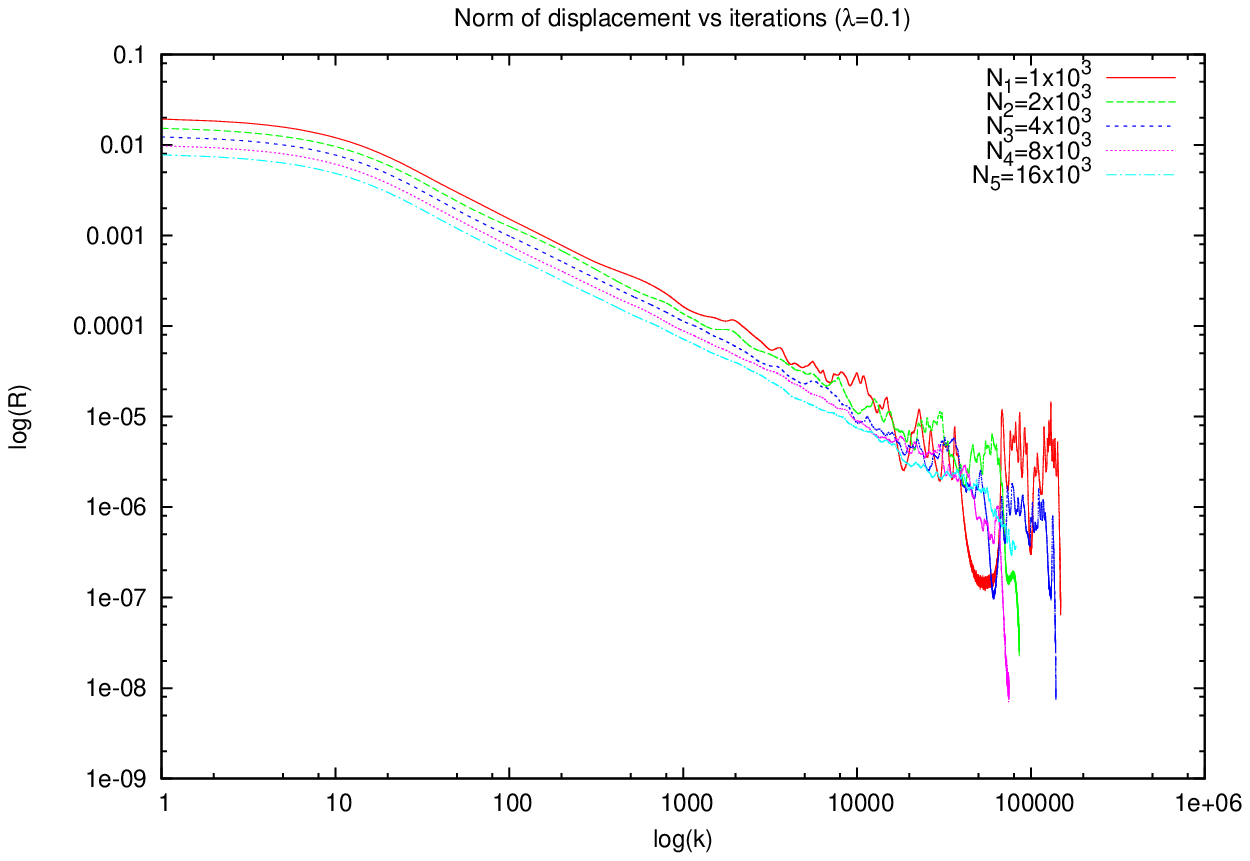} \\      
            \includegraphics[height=0.45\textwidth,width=0.45\textwidth]{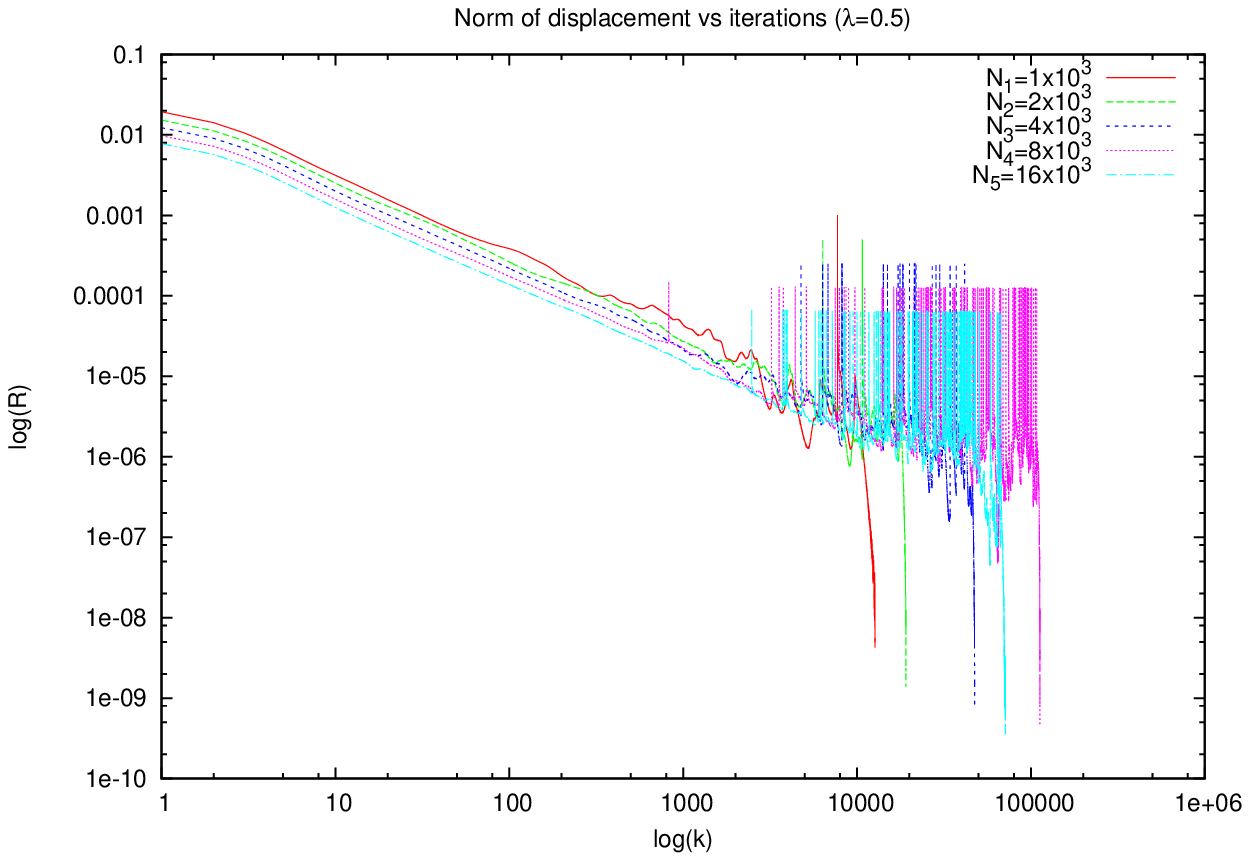}&
             \includegraphics[height=0.45\textwidth,width=0.45\textwidth]{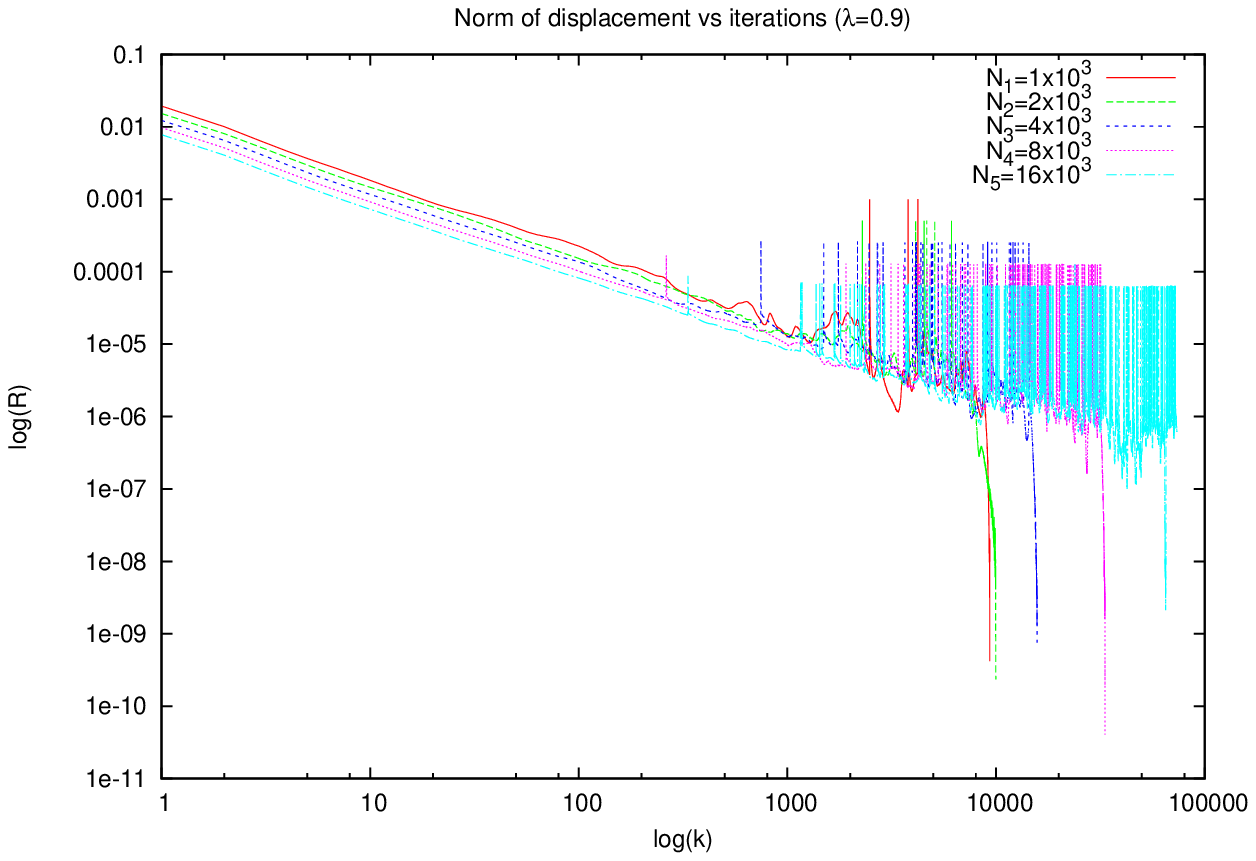}
  \end{tabular}
    \caption{We show in this graphics the generic behaviour of the norm of the displacement
    versus the iterations in the Lloyd's algorithm:
    {\bf Left top:} for few particles to hundreds of them we can see reach a
    stable configuration where all particles get the static conditions imposed over the algorithm, we found
    here the {\it cut iteration} $k_1=100, k_2=679, k_3=1065, k_4=1450, k_5=3489, k_6=7239$ for 
   $N_1=9,N_2=18,N_3=36, N_4=72, N_5=144, N_6=288$. For the following configurations we use  
   $N_1=1\times10^3,N_2=2\times10^3,N_3=4\times10^3, N_4=8\times10^3, N_5=16\times10^3$ and the
   the {\it cold iterations} are
    {\bf Right top:} $k_1=*, k_2=85872, k_3=*, k_4=75014, k_5=*$ , 
    {\bf Left down:}$k_1=12701, k_2=19132, k_3=47482, k_4=*, k_5=71428$, {\bf Right down:}
    $k_1=9330, k_2=9968, k_3=15742, k_4=33297, k_5=*$.
    The a steric (*) iterations are those configurations whose final states never reached in the
    until the checking process, observe how other configurations have reached to the cold state. }\label{fig:norms_lloyds}
\end{figure}

The relaxation for $N=9$ particles has symmetric final state, see Fig.\ref{fig:voro_show}. However
as we can see the final state when the number of particles is increased it is more difficult to 
define any kind of symmetry at its final state, however if we assign to each point the property of
having mass $m_i$ and if we use the SPH approximations the we are able to use it to test
the accuracy of the new distribution. Our claim is that for a CVT
the SPH average is better than for a simple random distribution.

To check the coldest iteration we define the norm of the average of the displacement vector in the $k$-esim iteration as 
\begin{equation}
R^{(k)}=\sum^N_{i=1} \sqrt{(x^{(k)}_c-x^{(k)}_i)^2+(y^{(k)}_c-y^{(k)}_i)^2+(z^{(k)}_c-z^{(k)}_i)^2}.
\end{equation} 

We have made some tests using $N=9,18,36,72,144,288$ number of particles  and we have found
the coldest iteration for $k_c=100,679,1065,1450,3489,7239$. This series of numerical experiments
for the Lloyd's algorithm can be appreciated in detail in Fig.\ref{fig:norms_lloyds}

In order to study the behaviour of the particle configurations in Lloyd's algorithm we have
tested it with $\lambda_1=0.1$, $\lambda_2=0.5$ and $\lambda_3=0.9$ for different number
or particles $N_1=1\times10^3$, $N_2=2 \times 10^3$, $N_3=4 \times 10^3$, $N_4=8 \times 10^3$
and $N_5=16 \times 10^3$. As we can see  in Fig.\ref{fig:norms_lloyds} the decay of $R^{(k)}$
behaves in similar way for different $\lambda$.  We can recognise from this graphics an
oscillating behaviour around the precision machine before some iteration number, we can distinguish this
region from the other because in log-log scale we appreciate a well define slope, we call the region
after the slope the {\it coldest region} because the particles after this iteration will be performing small 
movements. 

We postulate the following conjecture: {\it most of the configurations in the coldest region will posses a better
distribution than the original configuration and they own also almost the same $L_1$ norm for the error in SPH}. 
Then, even though the coldest iteration never be reached we can take it as a cold state.

For an arbitrary distribution in flat space time at the end of the Lloyd's algorithm
we get a distribution of particles such that
the configuration obeys the density profile $\rho$ but each tessellation posses different mass. It could work
for some other codes which are based in Voronoi tessellations, but in our case it is necessary to demand
that each one of the cells posses the same mass.

\section{Method}
\label{sec:method}

The method that we propose transforms a unitary density distribution on a cube of length one
into a configuration of particles with the profile $\rho(x^j)$ and coordinates $x^j$. Due to the method
of transformation of coordinates relates a uniform distribution with the new configuration, see equation (\ref{eq:inverse}),
we expect that once we obtain a more uniform distribution the transformation in $\rho(x^j)$ will
improve obtaining less dispersion.

The first part of this section explains how to generate the {\it initial seed}:uniform distribution
with constant density, specifically $\rho=1$ in a cube $(0,1) \times(0,1) \times (0,1)$.
The second part describes how to implement Lloyd's algorithm in this distribution in order to get
a more isotropic configuration. 

Roughly speaking, if we improve the uniform distribution, the new configuration obtained using the 
transformation of coordinates will be improved to. If it is right we will be able to see it on the SPH approximations
as a consequence.

\subsection{Initial seed:}

For a constant distribution $\rho(x,y,z)=1$ over a cube $(0,1) \times(0,1) \times (0,1)$ in flat space:
\begin{enumerate}
\item Determine the elements of the $ADM$ metric \footnote{For flat space time but in other coordinate system, like polar spherical coordinate,
we have $\eta_{ij}=diag(1,r^2,\sin{\theta}^2)$}:
\begin{eqnarray}
\alpha=1, \, \beta^i=0, \, \eta_{ij}=\delta_{ij},
\end{eqnarray}
\item Find the integral (\ref{eq:mo_rel}) 
\begin{equation}
\int^1_0dx\int^1_0dy\int^1_0dz=1,
\end{equation}
\noindent this is a trivial integration and will simplify the operations
\item Deduce the cumulative functions for each one of the coordinates 
\begin{eqnarray}
F_x(\omega_x)=\int^{\omega_x}_0dx=y_x, \,
F_y(\omega_y)=\int^{\omega_y}_0dx=y_y, \,
F_z(\omega_z)=\int^{\omega_x}_0dx=y_z.
\end{eqnarray}
\item Find the inverse function, in this case trivial.
\begin{eqnarray}
\omega_x=x, \,
\omega_y=y, \,
\omega_z=z.
\end{eqnarray}
\noindent Here $x,y,z \in (0,1)$ are uniform random variables.
\item Repeating this process $N$ times we obtain $x_i,y_i,z_i$ and
using the coordinate transformation we obtain
$\omega_{x,i},\omega_{y,i},\omega_{z,i}$ with $i=1,\dots,N$ such that this obeys the desired configuration
(in this example a cube of constant density). However we have to notice we
are going to use Lloyd's algorithm to relax using the $x,y,z$ set of points.
\end{enumerate}

\subsection{Lloyd's Algorithm implementation}
As we explain before, this algorithm is applied directly on the uniform distribution in a cubical region of dimension one, $A^{(0)}$.
After $k_c$ iterations if the relaxed configuration is reached we transform this $A^{(k_c)}$ using the transformation coordinates described below.
The Lloyds algorithm is described in the following steps:

\begin{enumerate}
\item Set  the random distribution $A^{(0)}$ for a cubical uniform distribution.
\item We compute the Voronoi Tessellations using {\it Voro++}\cite{chrisvoro++}. 
This very useful library give us all the important features of the 
Voronoi Tessellations, we are using here only the position of the centroid of each cell.
\item Use the information of the centroid of the $VT$ in the k-esim iteration in order to move
each one of the $N$ particles to the centroid using 
\begin{equation}
\vec{x}^{(k+1)}_i=\vec{x}^{(k)}_i+ \lambda (\vec{x}^{(k),c}_i-\vec{x}^{(k)}_i),
\end{equation}
\noindent where $\vec{x}^{(k)}_i=(x_i,y_i,z_i)$ is the current position of the generator of the
VT (at $k=0$ is the initial seed, not the transformed),
$\vec{x}^{(k),c}$ are the coordinates of the centroid of the current VT and $\lambda$ is a factor
which drives the approaching to the centroid of VT, we have proved with different values of $\lambda$.
\item Repeat this process until $\| \vec{x}^{(k+1)}-\vec{x}^k\| \leq TOL$, where $TOL$ is a parameter
related to the machine precision. 
\end{enumerate}

Then the final state of this process is a Centroidal Voronoi Tessellation of a uniform distribution
in three dimensions. Now we perform the coordinate transformation described in the first subsection
and the transformed configuration will be obtained.

\section{Applications to Physical Configurations}
\label{sec:applications}

In this section we present the analysis of the results analysing the $L_1$ \cite{VSpringel_arxiv} norm defined as

\begin{equation}
L_1:=\frac{1}{N_A} \sum^{N_A}_{i=1} |f(\vec{x}_i)-\left<f\right>_i|,
\end{equation}
\noindent here $N_A$ are the particles inside certain region $A$.
 
We will apply this norm for the special relativistic case (euclidean space) and for
the Tolman-Openheimer-Volkof solution describing a spherical symmetric configuration 
in static gravitational equilibrium (curved space).

\subsection{Special Relativistic Case}

We have set up a configuration of $N$ particles inside a cube, such that the velocity of 
each particle is equal to zero, the rest mass density is equal to one. This corresponds to
the coordinate transformation explained above.

\begin{figure}[h!]
  \centering
  \begin{tabular}{cc}
    \includegraphics[height=0.45\textwidth,width=0.5\textwidth]{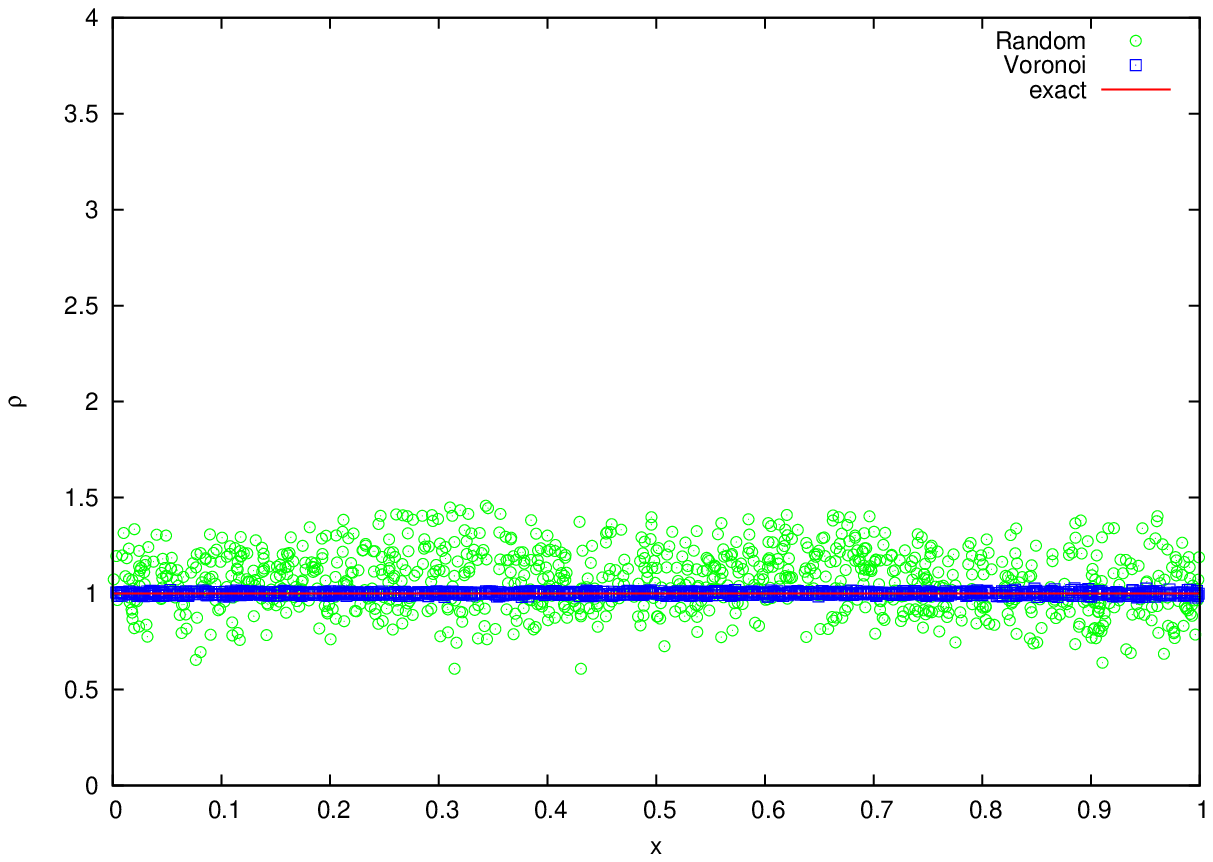}&
      \includegraphics[height=0.45\textwidth,width=0.5\textwidth]{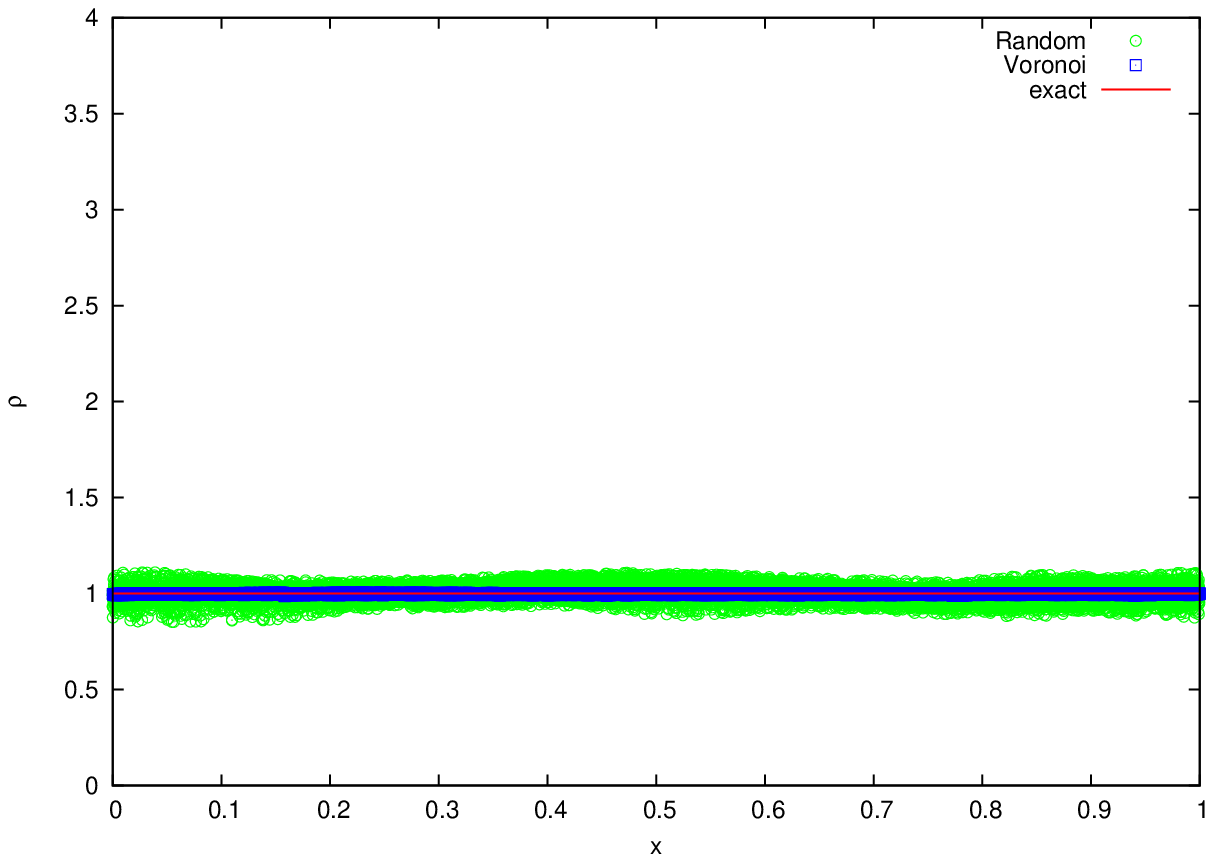} \\     
            \includegraphics[height=0.45\textwidth,width=0.5\textwidth]{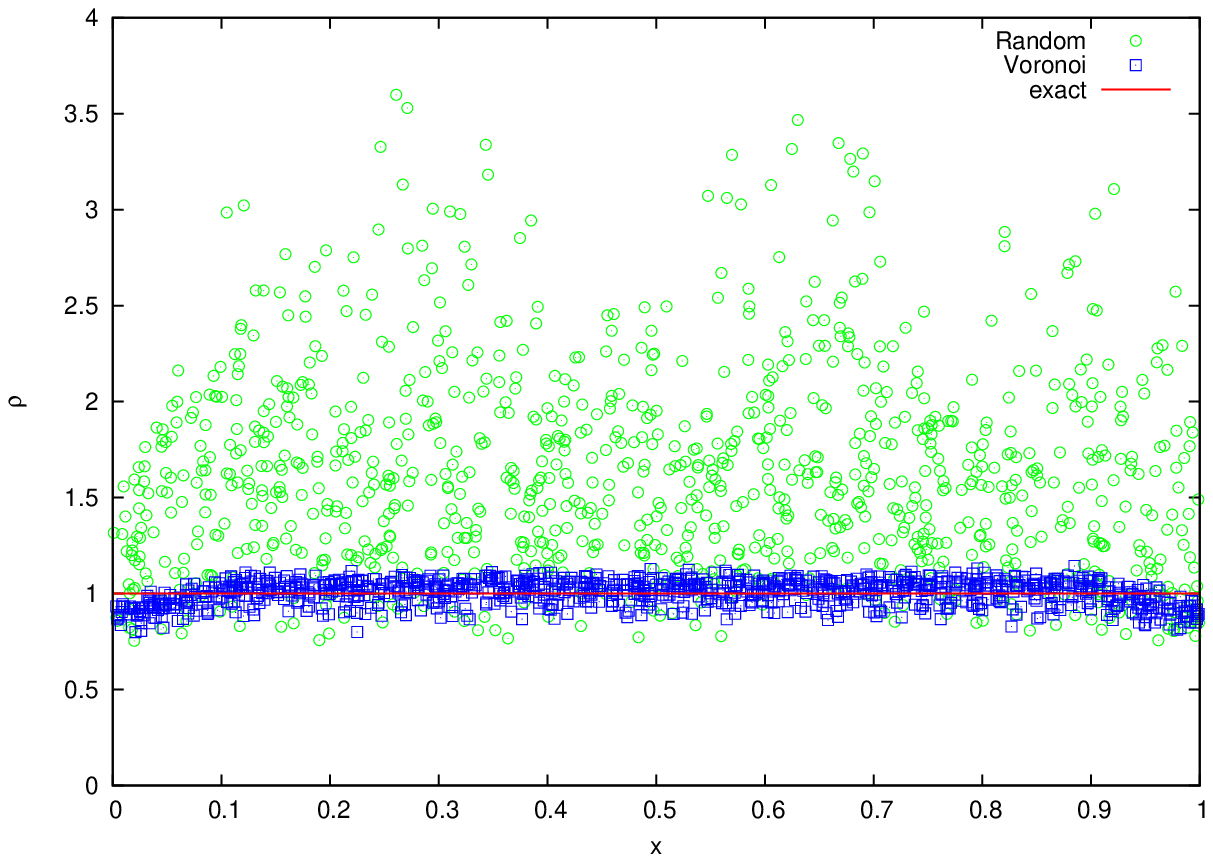}&
             \includegraphics[height=0.45\textwidth,width=0.5\textwidth]{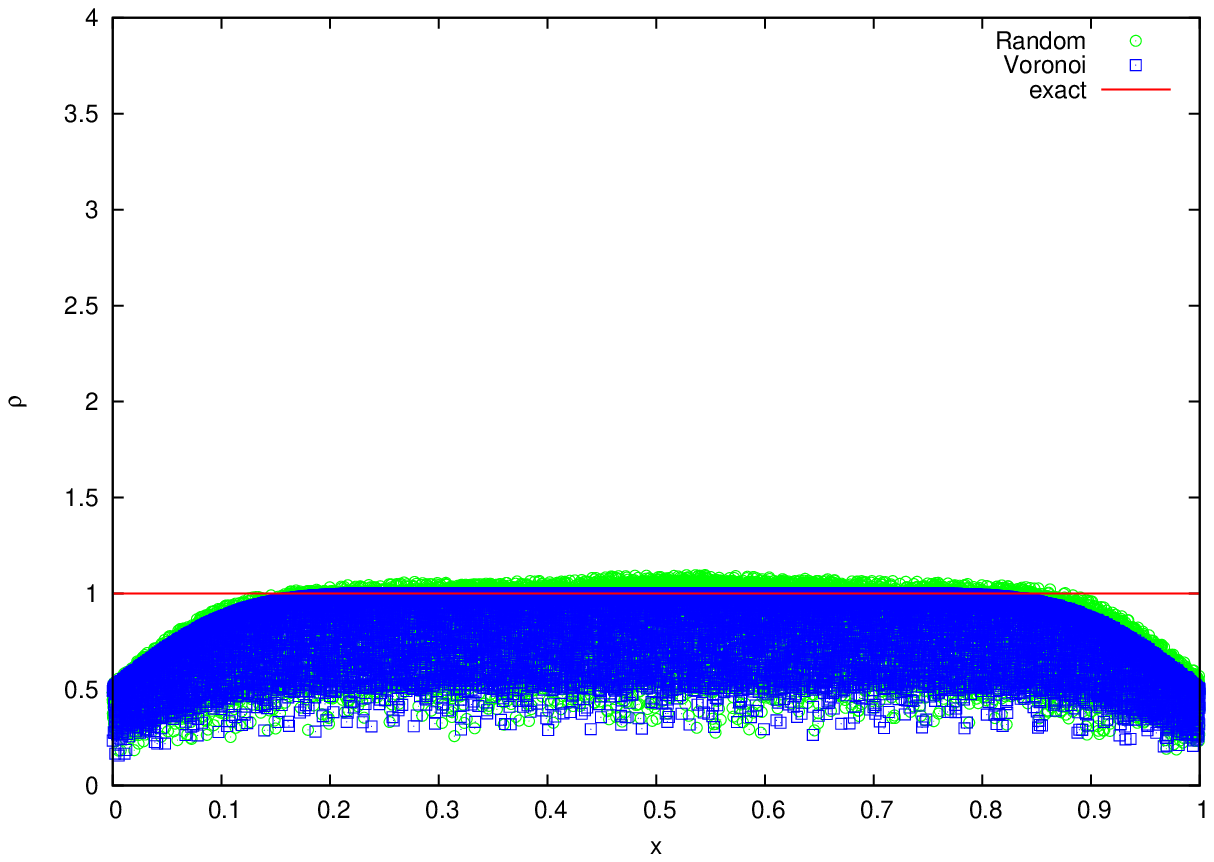}
  \end{tabular}
    \caption{In this graphics we show the free boundary averages for a random and Voronoi relaxed configuration,
    with periodic boundary conditions (up row) and for a free boundary conditions (down row). The decrement in the averages
    near the boundary are the result of the lack of particles in this regions. However the dispersion of the
    averages is less for Voronoi relaxed than for random configurations}\label{fig:voro_show_nfb}
\end{figure}

Using the relaxed configurations we have tested the SPH averages for several values
of particles, specifically in Fig.\ref{fig:voro_show_nfb} we can appreciate the corrective effects
of the relaxation configurations for $N=10^3$ and $N=16\times10^3$ over the random configurations.
In this figure also we can see  the effects of the boundary effects as we approach to $0$ and $1$
the average decays due to the lack of particles in the particles near to the boundary, we have tested with periodic
boundary effects and with free boundary conditions and we can appreciate a more accurate match and 
distinguish the effects that free boundary conditions produce.

\begin{figure}[h]
  \centering
  \begin{tabular}{c}
    \includegraphics[width=0.8\textwidth]{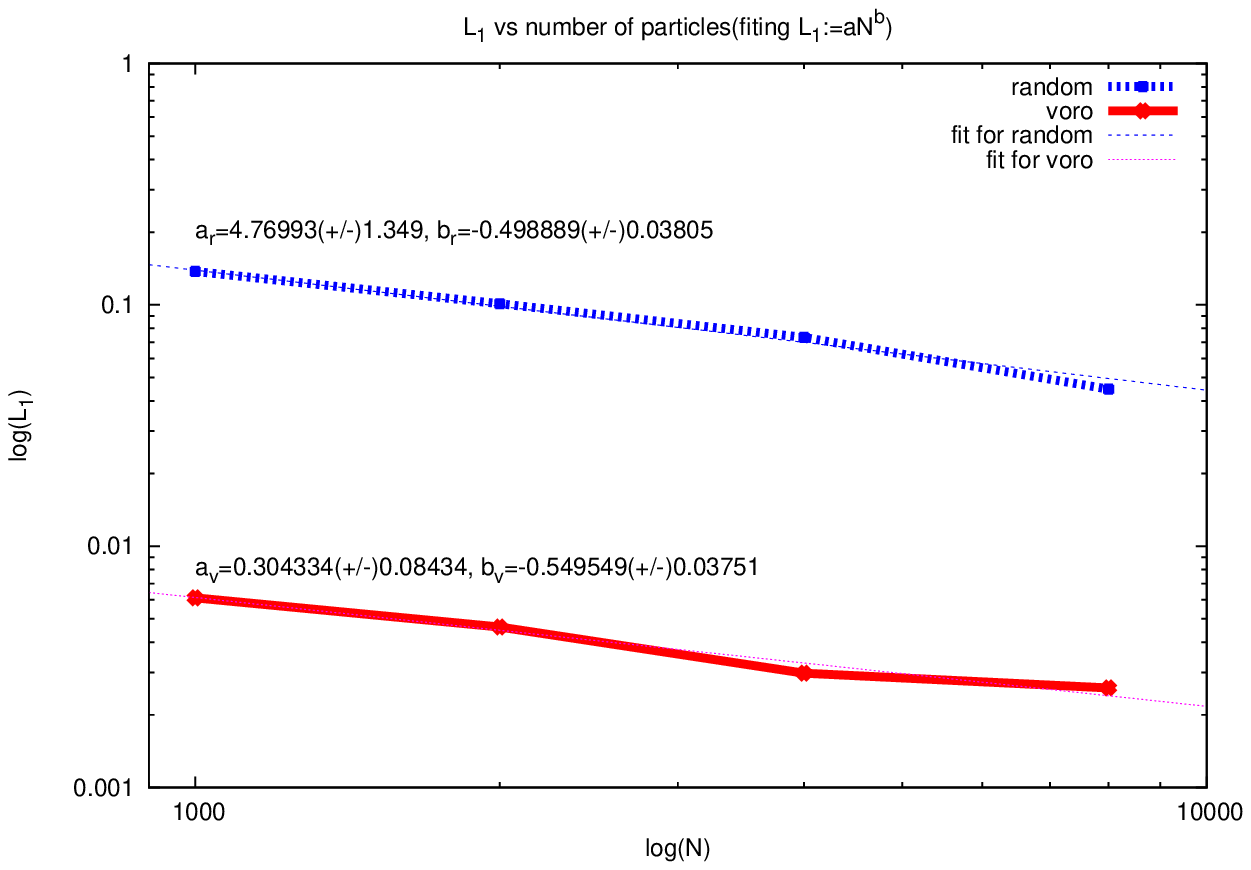}
    \end{tabular}
    \caption{Setting $N=10^3,2\times10^3,4\times10^3,8\times10^3$ we generate random and Voronoi relaxed 
    configurations. In this graphic we show the behaviour of the norm for both kind of configurations as well as
    the fitting line $L_1:=ax^b$. We can see that both of the configurations present the same $b$ approximately
    but the amplitude of $L_1$ is reduced in the Voronoi relaxed configurations in one order of magnitude approximately.}
   \label{fig:LvsN_special_rel}
\end{figure}

In order to check that the Voronoi relaxed configurations make an improvement on the average of particles
we take the $L_1$ norm defined before and compare the configurations for  $N=10^3,2\times10^3,4\times10^3,8\times10^3$ particles
for random and Voronoi relaxed configurations,  we found
that $L_1$ is better in Voronoi related configurations than for
the random configurations as we can see in Fig.\ref{fig:LvsN_special_rel}.

Trying to define a mathematical model for describing the behaviour of $L_1$ versus the $N$ we have used
$L_1(N):=aN^b$ and we found that random and Voronoi relaxed configurations posses almost the same
 exponent number $b_r=-0.498889$ and $b_v=-0.549549$ respectively. This is in agreement with  the order $O(1/\sqrt{N})$
 for the integration using Monte Carlo (remember SPH uses MC integration process \cite{Lucy77}), 
then we can assume that the Voronoi relaxation reduces the 
magnitude of the $L_1$ in one order approximately only by looking at the fraction $a_r/a_v$.

This analysis can not be done for the free boundary conditions because the lack of neighbours for the boundary
particles will corrupt the norm.

\subsection{TOV profile}

This case is the most important because it will prove how the present method for generating particle
distributions for SPH uses a coordinate transformation for inherit the well distributed properties of the uniform
distribution to a spherical distribution for general relativistic SPH codes.

In this applications we can see that the inversion rule 
$y^j=F^{-1}(X^j)$  is not trivial because the density is a numerical profile 
coming from the solution of Tolman-Openhaimer-Volkov (TOV)
equations system:
\begin{eqnarray}
\frac{dm}{dr} &=&4\pi r^2  \rho, \\
\frac{dP}{dr} &=& -\frac{\rho m}{r^2} \left( 1+ \frac{4\pi P r^3}{m}\right)\left( 1- \frac{2m}{r}\right)^{-1}\\
\frac{d\Phi}{dr} &=& -\frac{1}{\rho} \frac{dP}{dr} \left( 1+ \frac{P}{r}\right)^{-1}.
\end{eqnarray}
\noindent Here $m(r)$ is the total mass contained inside of an sphere of radius $r$, $\rho$ is the total mass density, $P$ is
the pressure and $\Phi(r)$ comes with the definition of the space-time where the fluid is living defined as
\begin{equation}
ds^2=-e^{2\Phi} dt^2 +e^{2\lambda} dr^2 +  r^2 d\Omega^2,
\end{equation}
\noindent where $d\Omega^2=d\theta^2+\sin^2 \theta d\phi^2,$ and $e^{2\lambda}:= \left( 1- {2m}/{r} \right)^{-1}$.
We are going to explain an alternative method for numerical density profiles.

\subsubsection{Coordinate transformation for TOV:}

\begin{enumerate}
\item Determine the $ADM$ components of the metric
\begin{eqnarray}
\alpha=e^{\Phi}, \, \beta^i=0, \, \eta_{ij}=diag(\left(1-2m/r\right)^{-1},r^2,\sin^2 \theta),
\end{eqnarray}
\noindent in the coordinate system $\{t,r,\theta,\phi\}$ where $t>0$, $0<r<R^*$, $0<\theta<\pi$ and $0<\phi<2\pi$.
\item Find the integral (\ref{eq:mo_rel}) 
\begin{equation}
M_o(R)=\int^{\pi}_0 \int^{2 \pi}_0 \int^R_0 \rho(r) \alpha u^t  \sqrt{\left(1-\frac{2M}{r} \right)^{-1} r^2 \sin^2 \theta}  dr d\theta d\phi,
\end{equation}
\noindent here $u^t=1$ and $\rho(r)$ and $\alpha(r)$ are given numerically by the integration of the system of equations mentioned above.
Then the integrand can be separated in one part depending only of $r$ and other of $\theta$, this gives
\begin{equation}
M_o(R)=\int^{2\pi}_0 d\phi  \int^{2 \pi}_0 \sin \theta d\theta \int^R_0 \rho(r) \frac{r}{\sqrt{1-\frac{2M}{r}}}  dr,
\end{equation}
\noindent as we can see from this integrals the radial part inherit all the curvature effects but leaves the mathematical problem as 
for a flat space-time with a modified density distribution.

\item The normalised integral help us to define the cumulative functions 
\begin{eqnarray}
F_r(r) &=&\int^{r}_0 \rho(r)  \frac{r}{\sqrt{1-\frac{2M}{r}}} dr, \\
F_\theta(\theta) &=& \frac{1}{2} \int^{\theta}_0 \sin \theta d\theta = \frac{1-\cos{\theta}}{2} \\
F_\phi(\phi) &=& \int^{\phi}_0\frac{1}{2\pi}d\phi= \frac{\phi}{2 \pi}.
\end{eqnarray}
\noindent As we can see $F_{\theta}$ and $F_{\phi}$ are the same as for flat space with spherical coordinates.
On the other hand, $F_r(r)$ will be treated
numerically because its integrand will be given by $f_r(r)=\rho(r)  {r}/{\left({1-{2M}/{r}}\right)} $  then its integral can be computed using
trapezoidal methods constructing a numerical grid $r_k=k \Delta r$ where $k=1,\dots,N_r$ and $\Delta r= {R^*}/{N_r}$ and the nodes
$f^k_r:=f_r(r_k)$. We have an array $(r_k,F^k)$ where $F^k$ is the numerical integral from $0$ to $r^k$.
\item The $\theta$ and $\phi$ coordinate transformations are invertible. For the numerical distribution we need to explain how to get
$r=F^{-1}_r(x)$: suppose the numerical profile given by
$(r_k, F^k)$ where $r_k \in (0,R^*)$ and $F^k \in (0,1)$(a discrete cumulative function where $k=1,\dots,N_r$).
We generate from a uniform distribution $x$ such that $0<x<1$, the following step is to search among the $F_k$ list the $j-esim$
element such that $F^j<x<F^{j+1}$. Then immediately we find the approximated value for $r$ using the average $r=\left({r_j+r_{j+1}}\right)/{2}$ or
using a linear interpolant between the two points $(r_j,F_j)$ and $(r_{j+1},F_{j+1})$.

\item Repeating the process described above $N$ times we obtain $r_i,\theta_i,\phi_i$ and
using the spherical coordinate transformations 
\begin{eqnarray}
x_i &=& r_i \sin \theta_i \cos \phi_i, \\
y_i &=& r_i \sin \theta_i \sin \phi_i, \\
z_i &=& r_i \cos \phi_i,
\end{eqnarray}
\noindent we obtain the spherical distribution for the TOV in cartesian coordinates.
\end{enumerate}

This process transforms an arbitrary amounts of points in an euclidean space $(X,Y,Z)$ (random variables) to a new one with the desired coordinates $(r,\theta,\phi)$,
and the last step consist in transform the $(r,\theta,\phi)$ coordinate system to a pseudo-cartesian coordinates $(x,y,z)$ in order
to be able to produce the SPH averages.

\begin{figure}[h!]
  \centering
  \begin{tabular}{cc}
    \includegraphics[height=0.45\textwidth,width=0.5\textwidth]{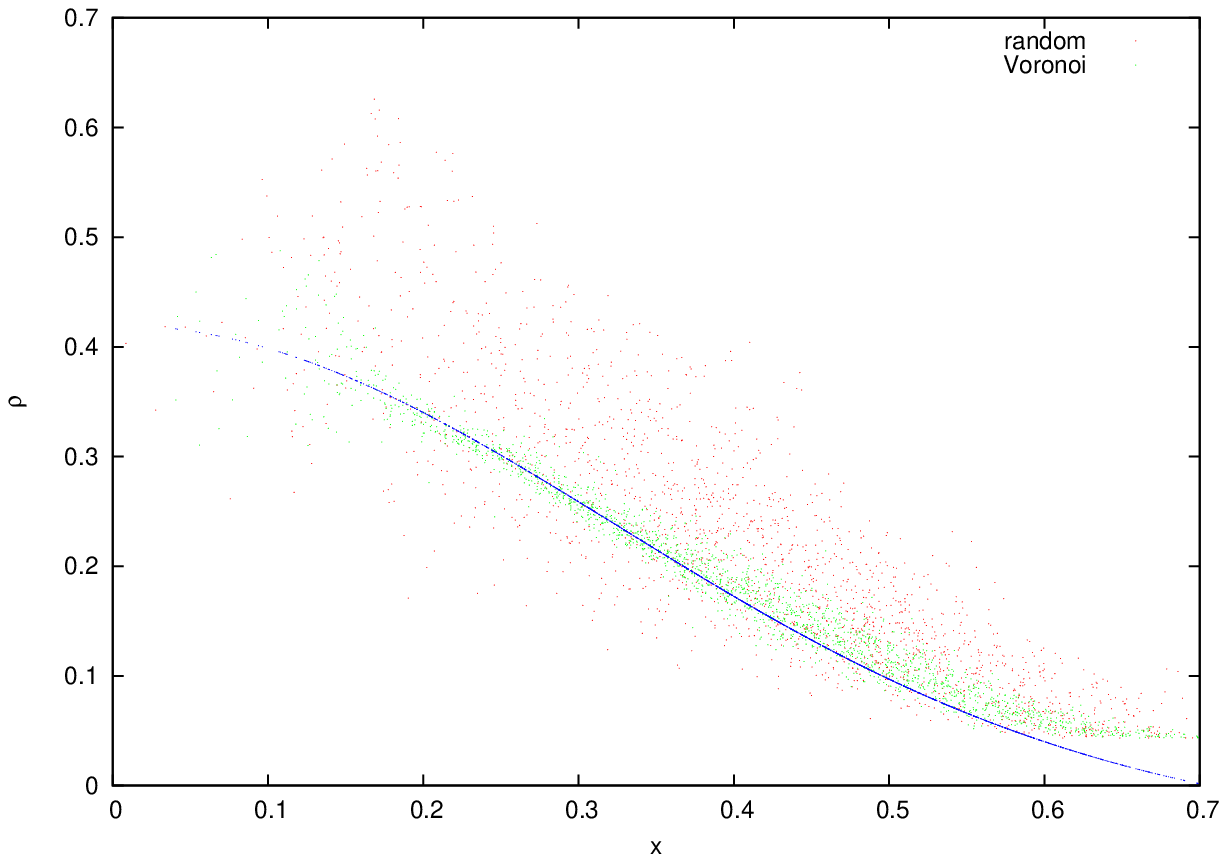}&
      \includegraphics[height=0.45\textwidth,width=0.5\textwidth]{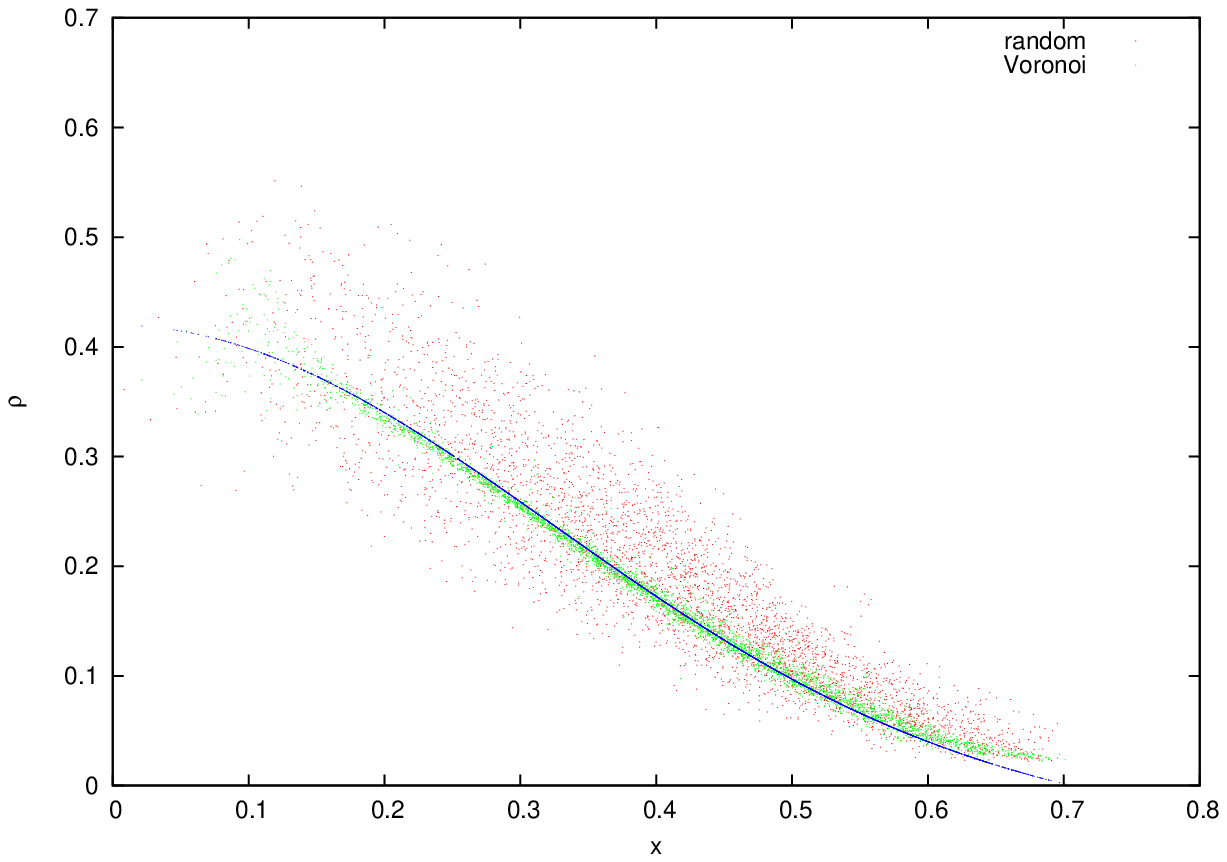} \\     
            \includegraphics[height=0.45\textwidth,width=0.5\textwidth]{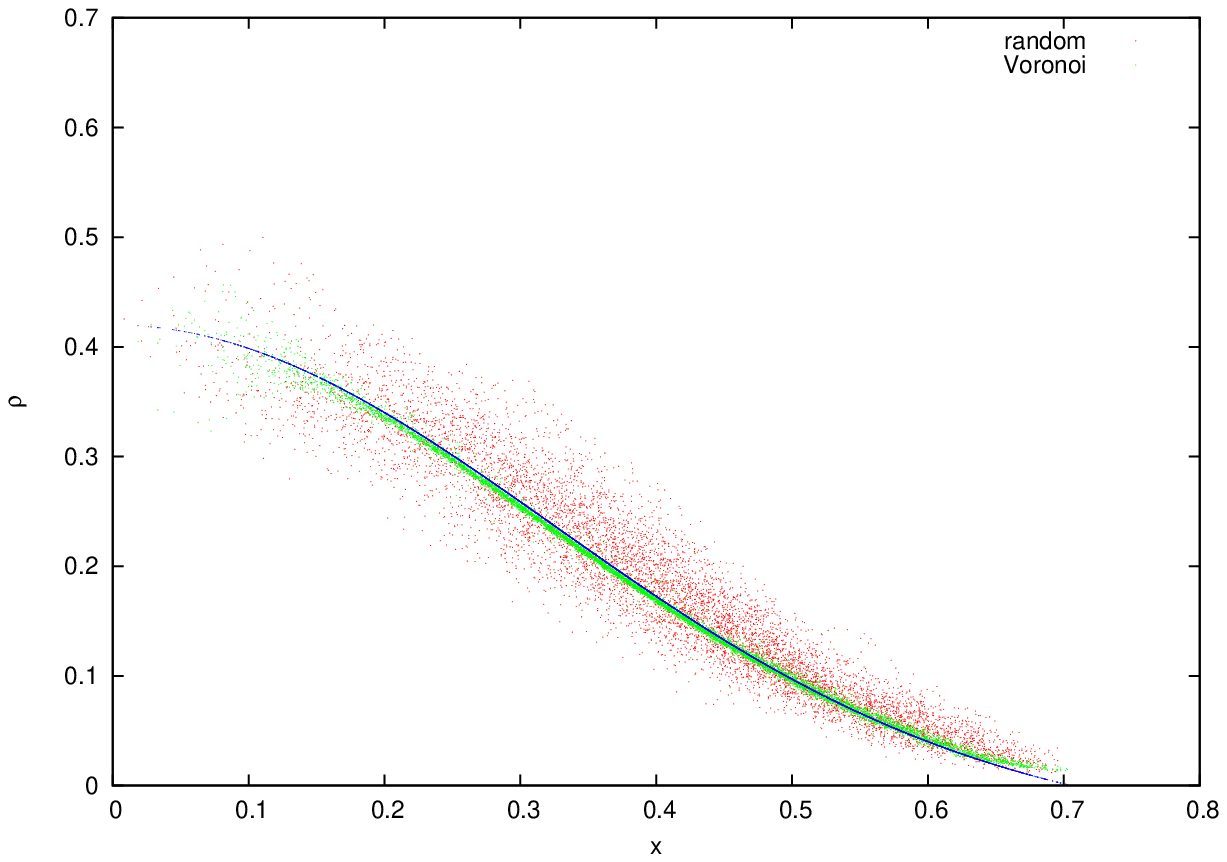}&
             \includegraphics[height=0.45\textwidth,width=0.5\textwidth]{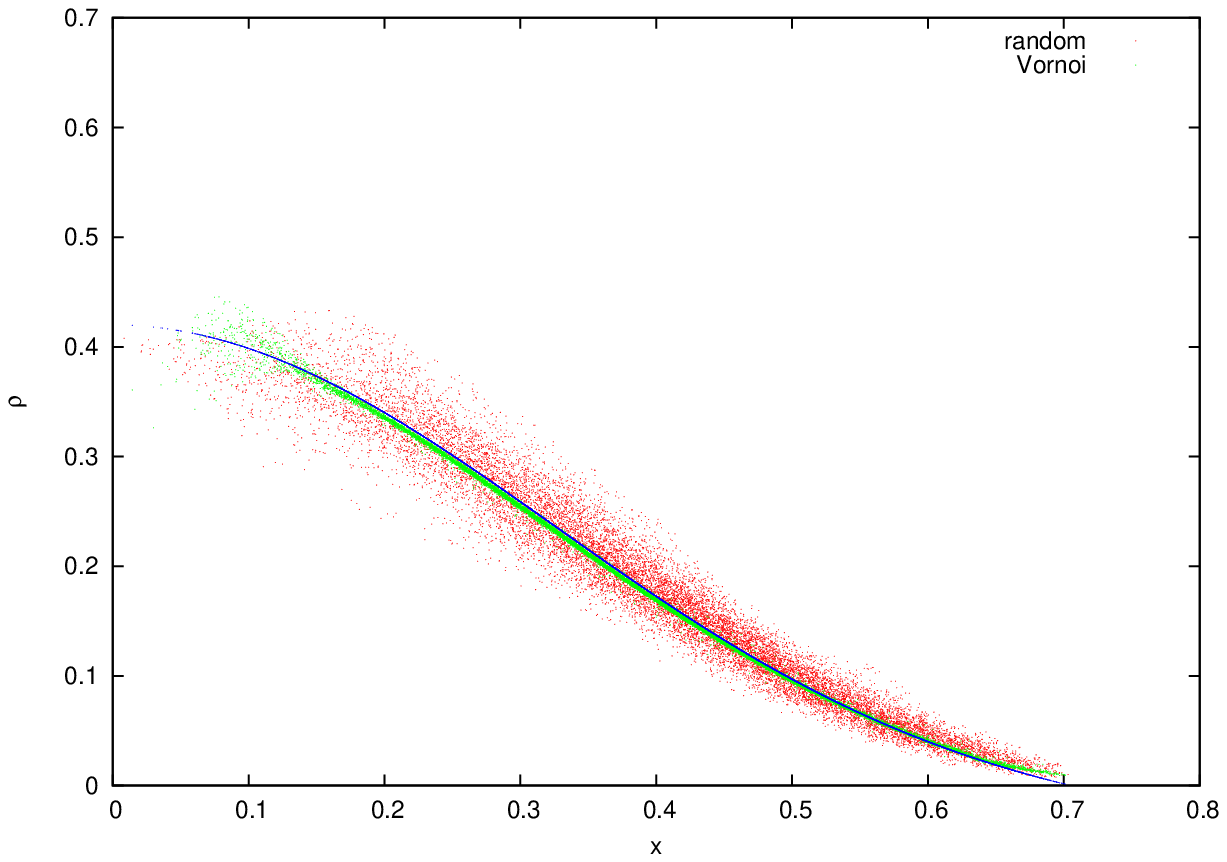}
  \end{tabular}
    \caption{We can appreciate in this graphic the effect of the transformation of coordinates on a Voronoi relaxed configuration, in the
    {\it top left} figure we have $N=2\times10^3$, in {\it top right} we use $N=4\times10^3$ particles, in {\it bottom left} $N=8\times10^3$ and
    in {\it bottom right} we have $N=16\times10^3$. As we can see even for an small amount of particles the corrective effects on the SPH
    average is easily appreciated, for $N=16\times10^3$ we can not distinguish between the analytic density profile an the Voronoi relaxed configuration
    transformed profile}.\label{fig:tovs}
\end{figure}

We made some tests for $N=\left\{1,2,4,8,16 \right\} \times 10^3$ particles with random
distributions and for Voronoi relaxed configurations, see Fig.\ref{fig:tovs}. 
As we can see the Voronoi relaxed configurations are much better than the random distributions, this improvement of the
SPH averages is a result of the more evenly distributed particles.

\section{Conclusions}

In this paper we have shown that in flat space-time the averages are better for Voronoi relaxed configurations than for random distributions,
we have shown using the $L_1$ norm that we need an enormous amount of particles in order to get the accuracy given by a CVT. Here we have
show also that the coordinate transformation inherit this properties to the configuration in the desired system of coordinates as we can see from
the SPH averages.

Although the Lloyd's algorithm can be used for obtaining particle distributions obeying an arbitrary density profile, 
it does not have the restriction that each particle have the same mass.
The volume of each particle can be approximated using the volume of each Voronoi cell contrary to the SPH averages 
in which the volume information is not required.

With this in mind we have described here a method which is dealing with particles with the same mass, compatible with the SPH simulations.
As a consequence as we increase the number of particles we have won in resolution faster than form random distributions, now we are in
a better position to describe accurately small regions of the fluid in the simulations.

As we can see this method can be used only under highly symmetric conditions for the distributions.





\bibliographystyle{model6-num-names}
\bibliography{my_database}





\end{document}